\documentclass[aps,prl,twocolumn,showpacs,superscriptaddress,10pt]{revtex4-1}
\pdfoutput=1

\usepackage{graphicx,mathrsfs}
\usepackage{amsmath,amssymb}
\usepackage{overpic}
\usepackage{xcolor}

\newcommand*{\vect}[1]{\mathbf{#1}}

\newcommand*{\ellk}{\ell_{\rm K}}
\newcommand*{\rv}{\vect{r}}

\begin{document}

\title{A model for melting of confined DNA} 
\author{E. Werner}
\affiliation{Department of Physics, University of Gothenburg, Sweden}
\author{M. Reiter-Schad}
\affiliation{Department of Astronomy and Theoretical Physics, Lund University, Sweden}
\author{T. Ambj\"ornsson}
\affiliation{Department of Astronomy and Theoretical Physics, Lund University, Sweden}
\author{B. Mehlig}
\affiliation{Department of Physics, University of Gothenburg, Sweden}
\date{\today}

\begin{abstract}
When DNA molecules are heated they denature. This occurs locally so that loops of molten single DNA strands form, connected by intact double-stranded DNA pieces.  The properties of this \lq melting\rq{} transition have been intensively investigated. Recently there has been a surge of interest in this question, caused by experiments determining the properties of partially bound DNA confined to nanochannels.  But how does such confinement affect the melting transition?  To answer this question we introduce, and solve a model predicting how confinement affects the melting transition for a simple model system by first disregarding the effect of self-avoidance. We find that the transition is smoother for narrower channels. By means of Monte-Carlo simulations we then show that a model incorporating self-avoidance shows qualitatively the same behaviour and that the effect of confinement is stronger than in the ideal case. 
\end{abstract} 

\maketitle
\begin{figure*}
\includegraphics{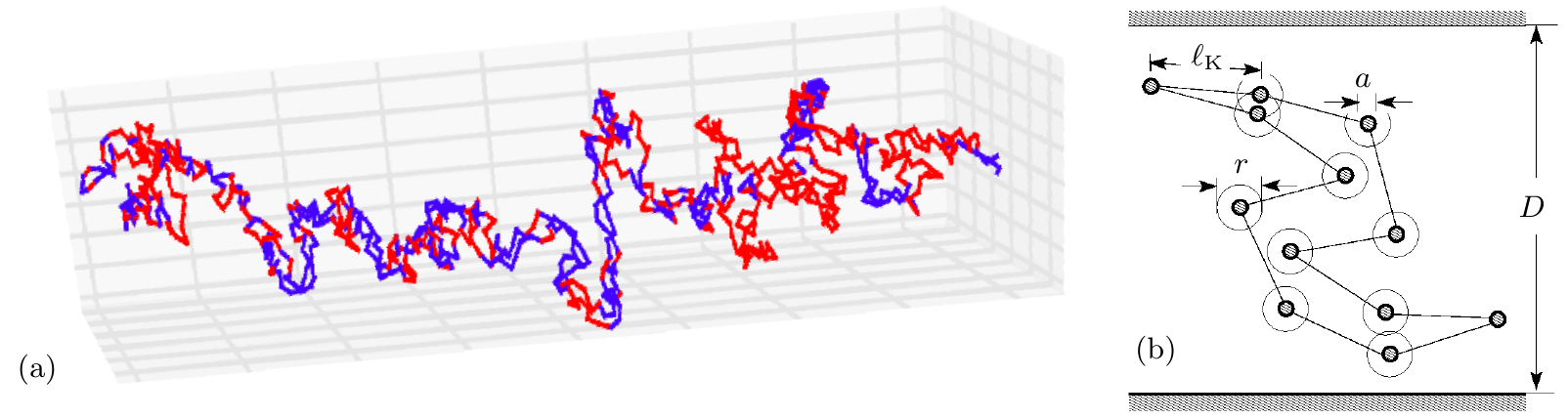}
\caption{  
\label{fig:modelSketch} (a) Conformation of a partially molten DNA molecule
confined to a channel of width $D=14 \ellk$. Shown is a snapshot
obtained from computer simulations of the model described in the text. 
The parameters of the simulation are 
$N=400$, $r=\ellk$, $a=0.63\ellk$, $E_{\rm b}=2.4$. The beads of diameter $a$ 
are not drawn.
Colour code: Molten parts (loops) are red, Intact parts are blue.
(b) Schematic illustration of the three-dimensional model of a confined double-
stranded chain that is used in the Monte-Carlo simulations, see text. Note that
the two strands are clamped at the ends.}
\end{figure*}

Thermal denaturation of DNA molecules has been studied intensively in the past
decades \cite{poland1966,frank-kamenetskii2014,santalucia1998,wartell1985,blake1999,
peyrard1989}. This
process, also referred to as \lq melting\rq{}, describes how the two strands
of the double-stranded helix separate when the hydrogen bonds holding together the 
base pairs disintegrate. When this happens at not too high temperatures,
loops or DNA \lq bubbles\rq{} of different sizes form. 
The fundamental question is to
determine which factors, apart from temperature, influence the shape of the
loop-size distribution, and how this distribution varies as the temperature is
changed.  A first step is to analyse a simple summary statistic -- the average
fraction of unbroken bonds. This binding probability is the order parameter
of the melting transition.

There are two important factors determining the temperature dependence of
this order parameter. First, {\tt G}-{\tt C} bonds are stronger than {\tt A}-{\tt T} 
bonds.
Since {\tt A}-{\tt T}-rich regions therefore melt more
easily \cite{wartell1985}, DNA sequence influences the melting transition. 
Second, entropy plays an important role. Loops and bound DNA strands
differ in how the number of spatial configurations depends on the number of
base pairs, and this difference determines the properties of the melting transition. 

Recently there has been great interest in DNA melting, stimulated by
the possibility to optically study sequence-specific local melting of DNA molecules
confined to nanochannels, see Ref.~\cite{reisner2012a} for a review. The aim
of these studies is to experimentally distinguish {\tt G}-{\tt C}-rich regions
from {\tt A}-{\tt T}-rich regions, obtaining sequence-dependent \lq bar
codes\rq{} representing given DNA sequences. Potential applications of this
method include fast species and strain identification \cite{reisner2012a}. In
the experiments described in Refs.~\cite{reisner2010,reisner2012a} DNA is confined to a
channel that is significantly smaller than the radius of gyration of the
unconfined DNA molecule.  In this case confinement substantially affects the
conformational fluctuations of the DNA molecule. 
This indicates that confinement must have
a strong effect upon the melting transition. For very short DNA molecules (15
base-pairs), the effect of confining the molecules to channels with a size of a
few nanometers has recently been studied by molecular dynamics simulations 
\cite{li2014}. Yet for longer molecules which exhibit loops of many different
sizes, it is not known how confinement influences the melting transition.

This motivated us to analyse an idealised model system describing the melting 
of a long double-stranded chain confined to a channel of width $D$.
In its simplest form the model disregards self-avoidance. In this case we can quantitatively explain the effect of confinement
upon the melting transition.
We find that the temperature dependence of the order parameter is smoother for  more
strongly confined chains. Our theory explains this effect by considering how 
confinement
affects the configurational entropy of single-stranded loops. 
By means of Monte-Carlo simulations we show that self-avoiding chains show a similar 
trend: confinement renders the melting transition smoother.
The effect is much stronger than in the ideal case (disregarding
self-avoidance).

The statistics of melting can be described by treating the DNA molecule as a
sequence of molten and intact sections. This method is often referred to as the Poland-Scheraga model \cite{poland1966}. 
Within this model, the statistics of DNA melting depend on the entropy
of a single-stranded loop. This entropy is
commonly assumed to be determined by the return probability of a random walk
representing a single DNA strand closing upon itself 
\cite{poland1966,fisher1966,kafri2000,kafri2002}.  
Our analysis shows that this is, in general, incorrect. We demonstrate that the
relevant quantity is instead the first return probability \cite{redner2001}. For
unconfined random walks in three dimensions this makes little difference since
these probabilities have the same power-law dependence upon loop size 
\cite{redner2001}. Yet it
is crucial for determining the entropy of large, confined loops. The entropy
of large loops in turn determines the order of the phase transition. By
analysing this entropy, we show that the order of the phase transition is
unchanged by confinement in a channel. This result is consistent with a study by
Causo {\em et al.} \cite{causo2002}, who show that for a lattice model of DNA melting,
the order of the phase transition is the same in one and three dimensions.
Yet as we discuss above, the melting
transition is strongly influenced by confinement. We infer that the order of 
the phase transition describes only a minute range of temperatures, whereas the
overall shape of the temperature dependence of the order parameter depends 
strongly on the value of first return probability for smaller loops.

{\em Model.}
We treat a highly idealised  model of double-stranded DNA, illustrated in 
Fig.~\ref{fig:modelSketch}.
Each strand consists of a freely jointed chain of $N$ spherical beads of diameter $a$
and excluded volume $v=4\pi a^3/3$, connected by ideal rods of
length $\ellk$.  The monomers of the first chain are labeled by $n=1,2,\ldots,N$.  
The monomers of the second chain are labeled by $n'$ (same range).  
The two chains are attached to each
other at both ends, as shown in Fig.~\ref{fig:modelSketch}(b).
For a given monomer $n$ and its partner $n'=n$ (for $1 < n=n' < N$) we test
whether the distance
between the two monomers is less than $r$. If so the two monomers
are considered bound, and a negative binding energy $-E_{\rm b}$ (in units such that 
$k_{\rm B}T = 1$) is
assigned to this pair. 
Note that binding is only allowed between
monomers corresponding to different chains, and only for $n=n'$.

{\em Effect of confinement on melting}. The statistics of melting is determined by
the difference in free energy between a bound section of $m$
base pairs and a molten section of the same size, bracketed by closed
base pairs \cite{kafri2000}. The bound section has a higher energy,
but also a higher entropy, and the probability of melting is
determined by the interplay of the two. Within our
model the increase in energy upon melting is given by the
number of broken bonds between the bases, 
proportional to $m$. The increase in entropy, on
the other hand, has a more complicated dependence upon $m$. To figure
out this dependence it is helpful to consider the two chains of
the DNA as independent chains, joined at one end. For the chains  
to remain bound they must move in
lockstep. The molten state is less restrictive, allowing any
configuration of the two strands that  fulfill two conditions. First,
the two strands must together from a closed loop, thus they must end
at the same position. Second, the loop must not close prematurely, as
that would not correspond to a single molten region of size $m$, but
rather two or more smaller molten regions. 
The second condition has not been considered in previous studies of melting
\cite{poland1966,fisher1966,kafri2000,kafri2002}. 
From this condition it follows that the entropy of a loop is not 
given by the return probability of a random walk, but by its first return 
probability.

\begin{figure}[t]
\includegraphics{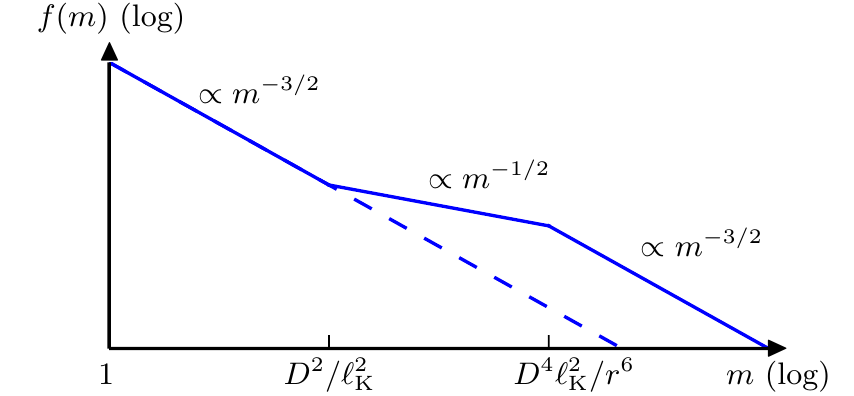}
\caption{\label{fig:firstReturnSketch}
A sketch of the first return probability $f(m)$. Dashed line: $f(m)$ for an 
unconfined DNA molecule. Solid line: $f(m)$ for ideal DNA confined to a square 
channel of size $D\gg \ellk$.}
\end{figure}

The first step in deriving a theory for melting is to determine the melting entropy 
$\Delta S(m)/k_{\rm B}$ 
by counting the number of configurations $\Omega_{\rm b}$ and $\Omega_{\rm u}$
satisfying the conditions of the bound and molten state, respectively.
For the ideal case of our model ($a=0$, no self-avoidance) we proceed as follows.
For the bound state, we estimate $\Omega_{\rm b} \approx c^{2m} p^m K(m)$.
Here $c$ is the number of ways in which a single Kuhn length segment can be oriented. 
For a spatially continuous model such as ours, it is in principle infinite. 
Imagine that space is finely discretised. Then $c$ is finite and we can show that $c$ drops 
out in the final result. 
Further 
$p = P(|\rv_k - \rv'_k| < r \,\,\mbox{\small and}\,\, |\rv_{k-1} - \rv'_{k-1}| < r)$
is the probability that monomer $k$ remains in lockstep, 
given that monomer $k-1$ is in lockstep, and $K(m)$ is the probability that a 
chain of length $m$ does not leave the channel.
This probability can be determined from the solution of a diffusion equation \cite{
casassa1967,werner2013}:
\begin{align}
\label{eqn:K_expression}
K(m) \approx \exp\left[-{\pi^2\ellk^2 m}/{(3 D^2)}\right]\,.
\end{align}
For the molten state, we find $\Omega_{\rm u} \approx c^{2m} K(m)^2 f(m).$
The function $f(m)$ is the probability that the two chains form a single closed loop, 
or alternatively the probability that the random walk 
performed by the separation $\rv_k-\rv'_k$ first returns to within a radius $r$
of the origin after $m$ steps. 
The function $f(m)$ is sketched in Fig.~\ref{fig:firstReturnSketch}.
For small values of $m$ the probability $f(m)$ is simply the first return probability 
for an unconfined three-dimensional random walk 
-- a short loop does not feel the presence of the walls. Assuming that $m\gg 1$, 
the first return probability scales as $f(m)\propto m^{-3/2}$ in this region 
\cite{redner2001}.
In the second and third regions the problem is essentially one-dimensional. 
Consider the component of the random walk in the channel direction. 
This component performs a one-dimensional random walk. Each time it returns to 
the origin, there is a probability $\kappa\propto r^3/(D^2\ellk)$ that the three
-dimensional random walk returns to within a radius $r$ of the origin 
\cite{werner2013,werner2014}. The probability of first return can therefore be 
mapped to the solved problem of finding the probability of absorption for a 
one-dimensional random walk with a partially absorbing sink at the origin
\cite{szabo1984}.
One finds that this probability scales as $f(m)\propto m^{-1/2}$ for small $m$, 
and as $f(m)\propto m^{-3/2}$ for large $m$. In summary, the function $f(m)$ has
the form sketched in Fig.~\ref{fig:firstReturnSketch}.
In the limit $r\to 0$, one need not distinguish between the probability of 
return and first return. We discuss this limit in the supplementary material.

Taking the above results together we have
\begin{align}
\label{eq:dS}
\Delta S/k_{\rm B} = \log\frac{\Omega_{\rm u}}{\Omega_{\rm b}} =\alpha \!+\! \gamma m 
\!-\! \frac{\pi^2\ellk^2 m}{3 D^2} + \log f(m).
\end{align}
Here $\alpha$ and $\gamma$ are undetermined constants which do not depend on whether 
the DNA is confined or not, $\alpha$ depends on the undetermined prefactors in 
our expressions for $\Omega_{\rm b}$ and $\Omega_{\rm u}$, and $\gamma = \log( 1/p)$ depends on 
the probability for a bound section to stay in lockstep.
The terms that are proportional to $m$ have the same effect as shifting the binding 
energy according 
to $\epsilon = E_{\rm b} -\gamma + {\pi^2\ellk^2 m}/({3 D^2})$.

Given Eq.~(\ref{eq:dS}) we can compute the 
order parameter, the probability $p_{\rm b}$ that a base pair is bound, 
at different binding energies. Let us first consider the qualitative behaviour 
of the melting curve in the thermodynamic limit ($N\to\infty$).
It is well known that the melting probability can be computed from a grand-canonical description (see for example 
Refs.~\cite{poland1966,fisher1966,kafri2000}), 
\begin{subequations}
\label{eqn:p_b_algorithm}
\begin{align}
p_{\rm b} &=  [1+{\rm e}^{\epsilon+\mu^*}\mathcal{Z}'_{\rm u}(\mu^*)]^{-1}\,, \\
\label{eqn:Z_u_definition}
\mathcal{Z}_{\rm u}(\mu^*) &=  
{\rm e}^\alpha\sum_{m=1}^\infty {\rm e}^ {m\mu^*} f(m)\,, \\
\label{eqn:muEquation}
1&={\rm e}^{\epsilon+\mu^*}[1+\mathcal{Z}_{\rm u}(\mu^*)]\, .
\end{align}
\end{subequations}
Here $\mathcal{Z}_{\rm u}(\mu)$ is the grand-canonical partition function of a molten section with chemical potential $\mu$,
and $\mathcal{Z}'_{\rm u}$ is its derivative with respect to $\mu$. 
Eq.~(\ref{eqn:muEquation}) specifies the value of the chemical potential necessary
to reach the thermodynamic limit.
If Eq.~(\ref{eqn:muEquation}) admits no solution for real $\mu^*$, 
we have  $p_{\rm b}=0$.

The effect of confinement upon the order parameter is determined by $f(m)$. The 
$D$-dependent correction to $\epsilon$ is small because
we assume that $\ellk \ll D$. 
At high values of $\epsilon$, Eq.~(\ref{eqn:muEquation}) yields 
$\mu^* \approx - \epsilon \ll 0$. For such large negative values of 
$\mu^*$, the sum in Eq.~(\ref{eqn:Z_u_definition}) converges rapidly.  The physical 
interpretation is that the formation
of large molten loops is very unlikely when $E_{\rm b}$ is large. The
order parameter is therefore determined by the behaviour of
$f(m)$ for small values of $m$. Yet here $f(m)$ is not influenced by
confinement as Fig.~\ref{fig:firstReturnSketch} shows.  In this region, then, 
confinement influences the melting curve only very slightly. 
In the opposite limit of negative values of $\epsilon$, the fact that $f(m)$ is 
significantly larger for the confined chain at large $m$  implies that 
Eq.~(\ref{eqn:muEquation}) admits solutions for a larger range of binding energies, and 
thus that $p_{\rm b}$ 
increases at lower energies for the confined chain, compared to the unconfined 
one. 
At intermediate binding energies, the binding probability depends sensitively on the 
exact shape of $f(m)$, and is therefore hard to predict. 
But we can conclude that the melting curve is sharper for the unconfined 
chain, and smoother for the confined one.

Note that although the shape of the melting curve is smoother for 
the confined chain, the order of the phase transition must remain unchanged, 
as the asymptotic scaling ($f(m) \propto m^{-3/2}$ as $m\to \infty$) is the same
for the confined and the unconfined chain. This 
shows that the order of the phase transition only determines the behaviour of the 
melting curve for a very small range of binding energies close to the point where $p_{\rm b}$
 becomes non-zero. Yet determining the location of this point and the full shape of 
the melting curve requires one to consider the behaviour of $f(m)$ for all $m$.

\begin{figure}[t]
\includegraphics{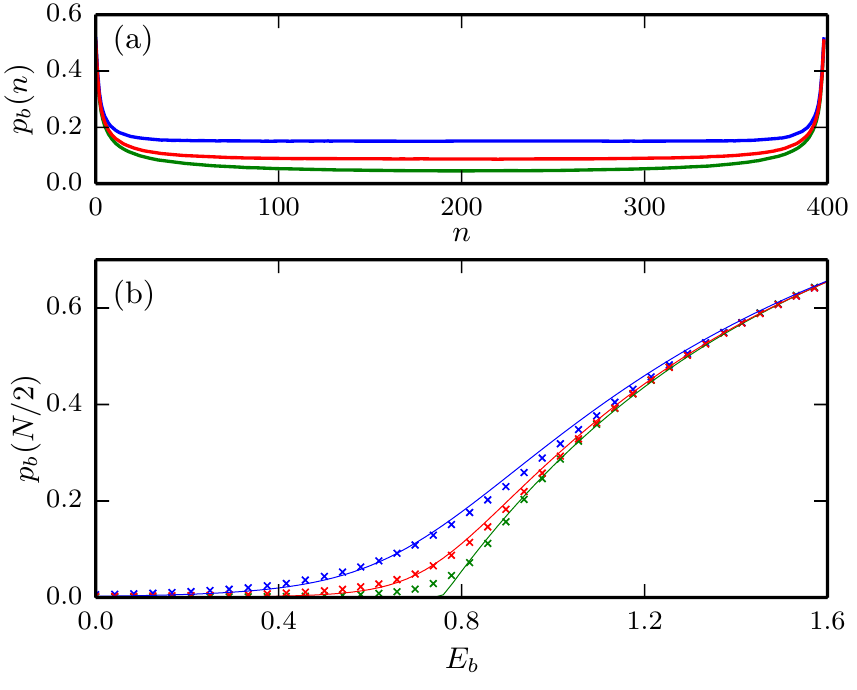}
\caption{\label{fig:bindingProbability_ideal}
Effect of confinement upon the melting transition, ideal case.
(a) Binding probability as a function of monomer number, for an ideal 
chain with binding energy $E_{\rm b}=0.78$, at different levels of confinement: 
$D/\ellk =8$ (blue), $D/\ellk =14$ (red), unconfined (green). $N=400$, 
$a=0$, $r=\ellk$
(b) Order parameter $p_{\rm b}(N/2)$ as a function of the binding energy, for different 
channel sizes as in panel (a). Solid lines: Fits to a solution of Eqs.~(3), see main text for
details.}
\end{figure}
 
{\em Simulation results.}
We performed computer simulations of our model using the Metropolis algorithm with
crankshaft trial updates \cite{yoon1995}. A snapshot from such simulations is shown 
in 
Fig.~\ref{fig:modelSketch}(a). We begin by comparing the results for the ideal case
 ($a=0$) to the theoretical predictions above.
Fig.~\ref{fig:bindingProbability_ideal}(a) shows the probability $p_{\rm b}(n)$ that a 
given base pair 
is closed, for an ideal chain with binding energy $E_{\rm b} = 0.78$, at different 
levels of confinement. 
For the upper two curves, this binding probability is independent of $n$ except close 
to the ends of the chain. This indicates that finite size effects are not important 
for these binding energies. The lower curve shows a case where the binding 
probability does not have a clearly 
plateau in the middle. In this case, then, finite size effects must 
result in a measured binding probability that is larger than it would be for an 
infinite chain. This effect is only present for small binding energies, and is 
largest 
for the unconfined chain. The simulations thus underestimate the sharpness of the 
unconfined melting curve.

Fig.~\ref{fig:bindingProbability_ideal}(b) shows how confinement influences the 
melting transition. We plot the probability that the middle base pair is closed, as a 
function of the binding energy 
$E_{\rm b}$. We find that confinement does not influence the melting probability at 
large $E_{\rm b}$, but also that increasing confinement leads to a smoother 
transition 
overall. These results are in perfect agreement with the theoretical expectations 
discussed above. 

We also show that the melting curve of Fig.~\ref{fig:bindingProbability_ideal}(b) 
can be well described by a numerical solution to Eq.~(\ref{eqn:p_b_algorithm}). 
For this numerical solution, we assume that $f(m) = (m + A)^{-3/2} + \lambda m^
{-1/2} \ellk^2/D^2$, consistent with the shape shown in Fig.~\ref{fig:firstReturnSketch}.
The empirical parameter $A$ is required to compensate for the fact that 
the diffusion approximation fails for very
short loops.
For $m > \lambda D^2/\ellk^2$, the function $f(m)$ crosses over to the scaling $
f(m)\propto m^{-1/2}$. $\lambda$ is another fitting parameter, which reflects
the fact that the exact location of the cross-over between $m^{-3/2}$ and $m^{-1/2}$ is not known. Since the
polymers that we have simulated are too short to exhibit the third region, we do
not include it in our numerical model.
Using this expression for $f(m)$, Eqs.~(\ref{eqn:p_b_algorithm}) can be evaluated
in Mathematica, yielding closed-form expressions for $p_{\rm b}(\mu^*)$ and
$\epsilon(\mu^*)$. Varying $\mu^*$ from $-\infty$ to 0 then allows one to trace
out the curve $p_{\rm b}(\epsilon)$. The parameters $\alpha,A,\lambda$, and $\gamma$ were determined using
the Mathematica routine {\tt FindFit}. First, the values $\alpha=-0.24$, $\gamma=1.58$
and $A=0.97$ were found by fitting the unconfined curve. Then,
keeping these values fixed, the value $\lambda=2.99$ was found by fitting
the curve corresponding to $D=8\ellk$. Although there are four fitting
parameters, the fact that the same parameter values yield good fits at different
confinement strengths confirms that our analysis accurately describes the effect
of confinement on the melting transition. That the overlap is not exact is to be
expected, as the function $f(m)$ is only approximately correct.
Note also that whereas the theoretical curve for the unconfined chain shows a
kink at low bending energies, the simulated curve is smooth. This discrepancy is
caused by the finite size of the simulated system (see discussion above).
\begin{figure}[t]
\includegraphics{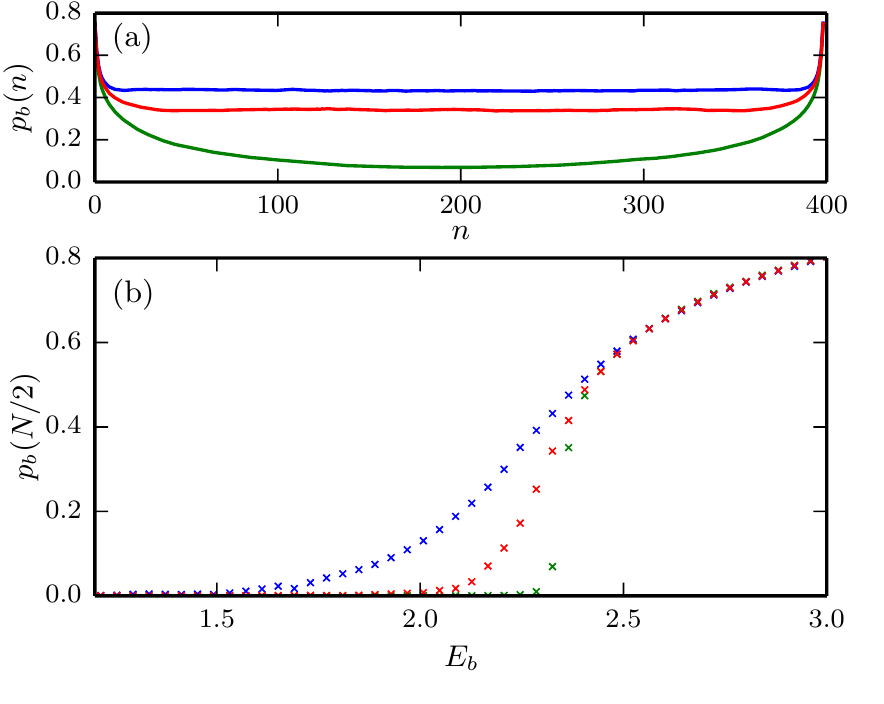}
\caption{\label{fig:bindingProbability_SA} 
Effect of confinement upon the melting transition, self-avoiding case.
(a) Binding probability as a function of monomer number, for a self-avoiding
chain with binding energy $E_{\rm b}=2.33$, at different levels of confinement: 
$D/\ellk =8$ (blue), $D/\ellk =14$ (red), unconfined (green). $N=400$, 
$a=0.63\ellk$, $r=\ellk$
(b) Order parameter $p_{\rm b}(N/2)$ as a function of the binding energy, for different 
channel sizes as in panel (a).}
\end{figure}

{\em Effect of self avoidance.} For the case of self-avoiding chains we do not
have theoretical predictions.  However, we have performed simulations of the model
described above with $a = 0.63\ellk$. The results are shown in
Fig.~\ref{fig:bindingProbability_SA}. Fig.~\ref{fig:bindingProbability_SA}(b)
shows that confinement has the same qualitative effect here as for the ideal
chain, in that it makes the transition smoother. Yet the effect is even more
noticeable in this case. Note that the for the case of the unconfined chain
(green curve), the transition is sharper than for an ideal chain [see
Fig.~\ref{fig:bindingProbability_ideal}(b)]. This is in agreement with
established theory for unconfined chains \cite{fisher1966,kafri2000}.
The fact that the lower curve in Fig.~\ref{fig:bindingProbability_ideal}(a) does
not exhibit a clear plateau indicates that for the unconfined chain, finite
size effects influence the melting probability at the transition. Also, compared to
the ideal case, much larger binding energies are required for binding to occur.
This is a consequence of the fact that it is much harder for a bound region
of the self-avoiding chain to remain in lockstep, as a large part of the binding
volume is inaccessible because of self-avoidance.

{\em Discussion.}
We have studied how the melting transition of DNA is modified by 
confinement. To this end, we have analysed a simple model by analytical calculations 
and Monte Carlo simulations. We show that confinement has a profound effect on the 
entropy of large molten sections, and that this results in a smoother melting 
transition than in the unconfined case. We show that the melting transition can 
be successfully analysed within the Poland-Scheraga model
 by calculating the entropy of a confined loop, and that 
this entropy is determined by the first return probability of a random walk. Our
theory is in quantitative agreement with simulation results for an ideal polymer
model. Simulations of a self-avoiding polymer show a similar, but even stronger 
effect of confinement.

The model that we study is clearly not a very realistic description of DNA. Yet the 
effect of confinement on the entropy of molten sections is insensitive to the details 
of the model. We therefore expect that channel confinement has 
a similar qualitative 
effect upon the melting transition, i.e. that the transition is smoother for confined 
DNA. The main difference is that for real DNA, the stiffness of dsDNA is much higher 
than that of ssDNA. Thus $K(m)$ [Eq.~(\ref{eqn:K_expression})] is different for a 
bound section than for a molten one. This has the effect of shifting the melting 
transition to lower temperatures, i.e. in the opposite direction compared to our 
model. We explore the consequences of our theory for real DNA in 
Ref.~\cite{michaela2015}.

{\em Acknowledgements}. BM acknowledges financial support by Vetenskapsr\aa{}det 
and by the
G\"oran Gustafsson Foundation for Research in Natural Sciences and Medicine.  
TA is grateful to the Swedish Research Council for funding (grant numbers 2009-2924 
and 2014-4305).
The numerical computations were performed using resources provided by C3SE and SNIC.

\end{document}


\title{Supplementary material for: \\A  model for melting of confined DNA} 
\author{E. Werner}
\affiliation{Department of Physics, University of Gothenburg, Sweden}
\author{M. Reiter-Schad}
\affiliation{Department of Astronomy and Theoretical Physics, Lund University, Sweden}
\author{T. Ambj\"ornsson}
\affiliation{Department of Astronomy and Theoretical Physics, Lund University, Sweden}
\author{B. Mehlig}
\affiliation{Department of Physics, University of Gothenburg, Sweden}
\date{\today}
\maketitle
In this supplementary material, we derive an explicit formula for the difference in
entropy $\Delta S (m)$ between a bound and a molten region, in the limit
$r\to 0$. We show that it agrees with the scaling in Eq.~(2) of the main text,
where $f(m)$ obeys the scaling of the first two regions of Fig.~2 (the third region,
where for large $m$ $f(m)\propto m^{-3/2}$, does not appear when $r\to 0$.)
We further note that while the exact expression of the formula depends on how
the region is assumed to be connected to the rest of the DNA, the scaling of
Eq.~(2) holds for all cases.

Denote by $G_N(\rv;\rn) d^3\rv$ the probability that a chain
started at $\rn$ survives the addition of $N$ monomers and ends up in the
volume $d^3\rv$ around $\rv$. If the channel size is much larger than the Kuhn
length of the chain, $G_N(\rv;\rn)$ obeys a diffusion equation \cite{casassa1967,
werner2013}
\begin{equation*}
 \frac{\partial G_N(\rv;\rn)}{\partial N} = \frac{\ellk^2}{6}\nabla_\rv^2 
G_N(\rv;\rn),
\end{equation*}
with absorbing boundary conditions: $G_N(\rv;\rn) = 0$ at the walls.
If the channel is square, with side length $D$, the solution can be found by separation of variables \cite{casassa1967,werner2013}
\begin{align*}
 G_{N}(\rv;\rn) &= G_N^\perp(x;x_0)G_N^\perp(y;y_0)G_N^\parallel(z;z_0),\mbox{ 
where} \\
 G_N^\parallel(z;z_0) &= \sqrt{3/(2\pi N \ellk^2)}\exp\lpa-\frac{3(z-z_0)^2}{2 
N \ellk^2}\rpa, \\ 
 G_N^\perp(x;x_0) &= \sum_{k=1}^ \infty\frac{2}{D}\sin\frac{k\pi 
x_0}{D}\sin\frac{k\pi x}{D}\exp\lpa-\frac{\ellk^2\pi^2k^2 
N}{6D^2}\rpa. 
\end{align*}
Since the $z$-direction is unconstrained, $G_N^\parallel(z;z_0)$ is simply the
probability distribution of a one-dimensional random walk. \\\hspace*{\fill}

 From these expressions we obtain the fraction of all chains which
``survive'' for $N$ links by integrating over all start
and end positions:
\begin{align}
\label{eqn:G_N_linear}
 \overline{G_N} &= \iint \frac{d^3\rn}{D^2} d^3\rv G_{N}(\rv;\rn)  \\
 &= \left[\sum_{k=0}^\infty 2 \left[\frac{2}{(2k+1) \pi}\right]^2 
\exp\lpa-\frac{\ellk^2\pi^2(2k+1)^2 N}{6D^2}\rpa\right]^2\!\!\!\!. \nonumber
\end{align}
This expression is approximately proportional to the number of configurations of
a bound segment under confinement, 
$\Omega_{\rm b}(m) \approx c^{2m} p^m \overline{G_{m}}$. 
See the main text for a discussion of the parameters $c,$ $p$.

Let us now turn to the other quantity of interest: $\Omega_{\rm u}(m)$, the 
number of possible configurations for a confined loop.   
The probability that a chain started at $\rv$
survives the addition of $N$ links and returns to within $d^3\rv$ of the
starting point is given by
\begin{align*}
 &G_{N}(\rv;\rv)d^3\rv = G_N^\perp(x;x)G_N^\perp(y;y)G_N^\parallel(z;z) d^3\rv
\\ & =\sqrt{3/(2\pi N \ellk^2)} \left[\sum_{k=1}^ 
\infty\frac{2}{D}\sin^2\frac{k\pi 
x}{D}\exp\!\!\lpa-\frac{\ellk^2\pi^2k^2 N}{6D^2}\rpa\right]  \\\times
&\left[\sum_{k=1}^ \infty\frac{2}{D}\sin^2\frac{k\pi y}{D}
 \exp\lpa-\frac{\ellk^2\pi^2k^2 N}{6D^2}\rpa\right]d^3\rv.
\end{align*}
Averaging the return probability over all starting positions $\rv$ yields
\begin{align}
\overline{G_{N}^\circ} & \!=\! \sqrt{3/(2\pi N \ellk^2)} D^{-2}
 \left[\sum_{k=1}^ \infty\exp\lpa\!-\frac{\ellk^2\pi^2k^2 
N}{6D^2}\rpa\right]^2\!\!\!. \label{eqn:returnProbability}
\end{align}
This quantity is proportional to the number of configurations
for a ring-polymer confined to a square nanochannel of
diameter $D$, $\Omega_{\rm u}(m) \propto c^{2m} r^3\overline{G_{2m}^\circ}$. 

The entropy difference between a bound and a molten region is given by $\Delta S(m)/k_B = \log
(\Omega_{\rm u}/\Omega_{\rm b})$.
We now show that this expression for $\Delta S$ agrees with Eq.~(2) of the main
text, with $f(m)$ scaling as the first two regions of Fig.~2 (recall that the third
region does not occur when $r\to 0$, which is the limit we treat here). 
Extracting the dominant exponential
terms from Eqs.~(\ref{eqn:G_N_linear}) and (\ref{eqn:returnProbability}) yields
\begin{align}
\label{eqn:dS_appendix}
\Delta S/k_{\rm B} &= \alpha + \gamma m 
- \frac{\pi^2\ellk^2 m}{3 D^2}  \\
&+ \log \left[\Omega_{\rm u}(2m) 
\exp\left(\frac{2\ellk^2\pi^2k^2 m}{3D^2}\right)\right] \nonumber\\
&- \log \left[\Omega_{\rm b}(m) 
\exp\left(\frac{\ellk^2\pi^2k^2 m}{3D^2}\right)\right]. \nonumber
\end{align}
As in Eq.~(2) of the main text, $\alpha$ and $\gamma$ are undetermined
constants. The last term in the expression is approximately constant, the
argument of the logarithm stays in the interval [1--1.53] for all $m$.
It remains to consider the expression $F(m)\equiv\Omega_{\rm u}(2m) 
\exp[2\ellk^2\pi^2k^2 m/(3D^2)]$. For large $m \gg D^2/\ellk^2$ only the
prefactor in Eq.~(\ref{eqn:returnProbability}) remains, yielding $F(m)\propto
m^{-1/2}$. In the opposite limit $m \ll D^2/\ellk^2$, Poisson resummation shows
that $F(m)\propto m^{-3/2}$. In summary, the function $F(m)$ exhibits the
same scaling as $f(m)$ in the limit $r\to 0$, and thus 
Eq.~(\ref{eqn:dS_appendix}) reproduces exactly Eq.~(2) of the main
text.

In the derivation of Eq.~(\ref{eqn:dS_appendix}), we averaged $G_n$ over
different positions, with all
positions in the channel weighted equally. While this procedure gives the
correct statistics for an isolated polymer, it is not quite correct for a region
in the interior of a larger polymer. Instead, the averaging must give more
weight to positions which are far from the channel walls \cite{werner2013}.
How the weighting is to be done depends on the state of the
surrounding polymer. If the region in question is in the interior of an
otherwise linear polymer the end position has a weight proportional to
$\sin(\pi x/D) \sin(\pi y/D)$ \cite{werner2013}. If, on the other hand, the region is bracketed by
large molten regions (loops), the weight is instead proportional to
$\sin^2(\pi x/D) \sin^2(\pi y/D)$.  It thus appears that  deriving an exact formula for 
$\Delta S(m)$ is not possible. However  all three averaging schemes
discussed here show the same general behaviour, given by Eq.~(2) and Fig.~2 of
the main text.

Note that in the limit $r\to 0$ the probability of return and the probability of first return are equal. Therefore, the third 
region in Fig.~2 of the main text cannot be obtained from the calculations 
described here. However, for small but non-zero values of $r$  it is easy to see that 
whenever the $z$-coordinates of 
two complementary monomers are close then there is a small, finite 
probability for binding to occur. As stated in the main text, this allows one to
map the problem of calculating the probability of first return to a solved 
one-dimensional problem \cite{szabo1984}, from which the existence of the
third region in Fig.~2 of the main text follows.